# Equilibrium Points and Periodic Orbits in the Vicinity of Asteroids with an Application to 216 Kleopatra

Yu Jiang

State Key Laboratory of Astronautic Dynamics, Xi'an Satellite Control Center, Xi'an 710043, China

Y. Jiang (✉) e-mail: jiangyu_xian_china@163.com (corresponding author)

**Abstract.** In this study, equilibrium points and periodic orbits in the potential field of asteroids are investigated. We present the linearized equations of motion relative to the equilibrium points and characteristic equations. We find that the distribution of characteristic multipliers of periodic orbits around the equilibrium point and the distribution of eigenvalues of the equilibrium point correspond to each other. The distribution of eigenvalues of the equilibrium point confirms the topology and the stability of periodic orbits around the equilibrium point.

## 1. Introduction

The interest for studying orbits around asteroids has received great attention. The study of orbits around asteroids is useful for two very important and current topics: navigation and orbital design of spacecraft near asteroids in sample-return missions (Bradley and Ocampo 2013) and the existence and exploration of small natural satellites or moonlets of irregular asteroids (Descamps 2010). Missions to asteroids may include the observation of shape and physical characteristics of asteroids, the soft-landing to asteroids, surface motion and sample-return, etc. The existence, dynamics and exploration of small natural satellites or moonlets of irregular asteroids have grown considerably recently. There are several binary asteroids and triple asteroids discovered in the solar system, such as the binary asteroid 243 Ida (Petit et al. 1997), the triple asteroid (47171) 1999 TC36 (Benecchi et al. 2010), etc. Besides, the study of orbits around asteroids is also useful for studying the ejection of dust particles from small celestial bodies (Oberc 1997) as well as the continuation of periodic orbits in the vicinity of asteroids (Scheeres et al. 1996; Yu and Baoyin 2012a, 2012b).

Orbits around rotating highly irregular-shaped celestial bodies include equilibria, periodic orbits, quasi-periodic orbits, and chaotic motions (Scheeres et al. 2012a; Jiang et al. 2014). Simple-shaped bodies or special potential fields are usually considered when studying dynamical behaviors, including equilibrium points, stability, and periodic orbits. These simple-shaped bodies include a logarithmic gravity field (Elipe and Riaguas 2003), a straight segment (Riaguas et al. 1999, 2001; Arribas and Elipe 2001; Elipe and Lara 2003; Romero et al 2004; Lindner et al. 2010; Najid et al. 2011), a triangular plate and a square plate (Blesa 2006), a circular ring (Lass and Blitzer 1983; Broucke and Elipe 2005; Najid et al. 2012), an annulus disk (Eckhardt and Pestaña 2002; Alberti and Vidal 2007; Fukushima 2010) as well as a cube (Chappell et al. 2012; Liu et al. 2011).

These explorations of dynamical behaviors around simple-shaped bodies can help us understand the dynamical behaviors around irregular-shaped celestial bodies.

Riaguas et al. (1999) pointed out that there exist bifurcations of periodic orbits in the gravity field of a massive straight segment with changing parameters. There are 4 equilibrium points in the gravity field of a logarithmic and a massive finite segment (Elipe and Riaguas 2003). Liu et al. (2011) found 8 equilibrium points in the gravity field of a cube, with 4 of them linearly stable, and the other 4 unstable.

Some highly irregular-shaped celestial bodies were analyzed to find periodic orbits around them, e.g., asteroids 216 Kleopatra (Yu and Baoyin 2012; Jiang et al. 2014), 433 Eros (Scheeres 2000), 4179 Toutatis (Scheeres et al. 1998; Scheeres 2012a), 4769 Castalia (Scheeres et al. 1996; Takahashi et al. 2013), as well as comet 67/P CG (Scheeres 2012a, b).

Here we will focus on the equilibrium points and periodic orbits in the potential field of asteroids. The linearized equations of motion relative to the equilibrium points and characteristic equations are discussed in Sect. 2. In Sect. 3 periodic orbits around equilibrium points are considered. We find that the distribution of characteristic multipliers of periodic orbits around the equilibrium point and the distribution of eigenvalues of equilibrium point correspond to each other; the distribution of eigenvalues of the equilibrium point confirms the topology and stability of periodic orbits around the equilibrium point.

**2. Motion Equations and Equilibrium Points**

Consider the motion of a massless test particle around a highly irregular-shaped celestial body. The Lagrangian of motion is

$$L = \frac{1}{2}(\dot{\mathbf{r}} + \boldsymbol{\omega} \times \mathbf{r})^2 - U(\mathbf{r}) \qquad (1)$$

where **r** is the body-fixed vector from the celestial body's center of mass to the particle, **ω** is the rotation angular velocity vector of the body relative to the inertial space, and $U(\mathbf{r})$ is the gravitational potential of the body. The Jacobian integral $H$ and the effective potential $V$ can be defined as (Scheeres et al. 1996; Yu and Baoyin 2012)

$$\begin{cases} H = \frac{1}{2}\dot{\mathbf{r}}\cdot\dot{\mathbf{r}} - \frac{1}{2}(\boldsymbol{\omega}\times\mathbf{r})(\boldsymbol{\omega}\times\mathbf{r}) + U(\mathbf{r}) \\ V(\mathbf{r}) = -\frac{1}{2}(\boldsymbol{\omega}\times\mathbf{r})(\boldsymbol{\omega}\times\mathbf{r}) + U(\mathbf{r}) \end{cases} \quad (2)$$

For a zero-velocity manifold for the particle $V(\mathbf{r}) = H$; for the forbidden region $V(\mathbf{r}) > H$, while for the allowable region $V(\mathbf{r}) \leq H$. Thus the equation of motion relative to the uniformly rotating body can be written as (Scheeres et al. 1996; Yu and Baoyin 2012)

$$\ddot{\mathbf{r}} + 2\boldsymbol{\omega}\times\dot{\mathbf{r}} + \frac{\partial V(\mathbf{r})}{\partial \mathbf{r}} = 0 \quad (3)$$

The linearized equations of motion relative to the equilibrium point can be expressed as

$$\begin{aligned} \ddot{\xi} + 2\omega_y\dot{\zeta} - 2\omega_z\dot{\eta} + V_{xx}\xi + V_{xy}\eta + V_{xz}\zeta &= 0 \\ \ddot{\eta} + 2\omega_z\dot{\xi} - 2\omega_x\dot{\zeta} + V_{xy}\xi + V_{yy}\eta + V_{yz}\zeta &= 0 \\ \ddot{\zeta} + 2\omega_x\dot{\eta} - 2\omega_y\dot{\xi} + V_{xz}\xi + V_{yz}\eta + V_{zz}\zeta &= 0 \end{aligned} \quad (4)$$

where $V_{\mathbf{rr}} \triangleq \begin{pmatrix} V_{xx} & V_{xy} & V_{xz} \\ V_{xy} & V_{yy} & V_{yz} \\ V_{xz} & V_{yz} & V_{zz} \end{pmatrix}_L$ is the Hessian matrix of $V(\mathbf{r})$. Denote $\lambda$ as the set of eigenvalues of the equilibrium points. Then the characteristic equation for the eigenvalues can be expressed as

$$\lambda^6 + \left(V_{xx} + V_{yy} + V_{zz} + 4\omega_x^2 + 4\omega_y^2 + 4\omega_z^2\right)\lambda^4 + \left(V_{xx}V_{yy} + V_{yy}V_{zz} + V_{zz}V_{xx} - V_{xy}^2 - V_{yz}^2 - V_{xz}^2 \right.$$
$$\left. + 8\omega_x\omega_z V_{zx} + 8\omega_y\omega_z V_{yz} + 8\omega_x\omega_y V_{xy} + 4\omega_x^2 V_{xx} + 4\omega_y^2 V_{yy} + 4\omega_z^2 V_{zz}\right)\lambda^2 \qquad . \quad (5)$$
$$+ \left(V_{xx}V_{yy}V_{zz} + 2V_{xy}V_{yz}V_{xz} - V_{xx}V_{yz}^2 - V_{yy}V_{xz}^2 - V_{zz}V_{xy}^2\right) = 0$$

If there is no eigenvalue that equals zero, the equilibrium point is non-degenerate (Jiang et al. 2014).

Denote $\omega$ as the modulus of the vector $\boldsymbol{\omega}$; then the unit vector $\mathbf{e}_z$ can be expressed by $\boldsymbol{\omega} = \omega \mathbf{e}_z$. Define the body-fixed frame through a set of orthonormal right-hand unit vectors $\mathbf{e}$

$$\mathbf{e} \equiv \begin{Bmatrix} \mathbf{e}_x \\ \mathbf{e}_y \\ \mathbf{e}_z \end{Bmatrix}. \qquad (6)$$

Then the characteristic equation (Eq. (5)) can be simplified to (Jiang et al. 2014)

$$\lambda^6 + \left(V_{xx} + V_{yy} + V_{zz} + 4\omega^2\right)\lambda^4 + \left(V_{xx}V_{yy} + V_{yy}V_{zz} + V_{zz}V_{xx} - V_{xy}^2 - V_{yz}^2 - V_{xz}^2 + 4\omega^2 V_{zz}\right)\lambda^2$$
$$+ \left(V_{xx}V_{yy}V_{zz} + 2V_{xy}V_{yz}V_{xz} - V_{xx}V_{yz}^2 - V_{yy}V_{xz}^2 - V_{zz}V_{xy}^2\right) = 0 \qquad , \quad (7)$$

The asteroid 216 Kleopatra has a peculiar shape that looks like a dog bone; the overall dimensions are $217 \times 94 \times 81$ km (Ostro et al. 2000), and the estimated bulk density is $3.6$ $\mathrm{g \cdot cm^{-3}}$, the rotational period is 5.385 h (Descamps et al. 2011). There are two moonlets near 216 Kleopatra, Alexhelios (S/2008 (216) 1) and Cleoselene (S/2008 (216) 2) (Descamps et al. 2011). These two moonlets are about 5 km and 3 km in diameter, respectively. In addition, there are 9 such triple asteroid systems discovered: 45 Eugenia (Merline et al. 1999; Marchis et al. 2010), 87 Sylvia (Marchis et al. 2005), 93 Minerva (Marchis et al. 2011), 216 Kleopatra, 3749 Balam (Vokrouhlický et al. 2009), 47171 1999TC36 (Stansberry et al. 2006; Benecchi et al. 2010), 136108 Haumea (Pinilla-Alonso et al. 2009), 136617 1994CC (Brozović et al.

2011; Brozović et al. 2010), and 153591 2001SN263 (Fang et al. 2011; Araujo et al. 2012). Seven of them are large size triple asteroids: 45 Eugenia, 87 Sylvia, 93 Minerva, 216 Kleopatra, 136108 Haumea, 136617 1994CC, and 153591 2001SN263. Among these large size triple asteroids, only 216 Kleopatra has a radar shape model (Neese 2004). So we choose 216 Kleopatra for our study of the dynamics around an asteroid. The asteroid 216 Kleopatra has a rotating principal axis (Ostro et al. 2000), and the z-axis of the body-fixed frame is taken as the principal axis. There are four equilibrium points outside the body of 216 Kleopatra (Yu and Baoyin 2012; Jiang et al. 2014). Figure 1 shows these four equilibrium points in the vicinity of 216 Kleopatra. The equilibrium points are denoted as E1, E2, E3, and E4. Positions of these equilibrium points are given in Table 1. E1 and E2 belong to Case 2 (see section 3.1 for a definition of Cases 1 through 5) while E3 and E4 belong to Case 5 (Jiang et al. 2014). The physical model of 216 Kleopatra that we use here was calculated from radar observations (Neese 2004). This polyhedral model has 2048 vertices and 4096 faces (Werner 1994; Werner and Scheeres 1997).

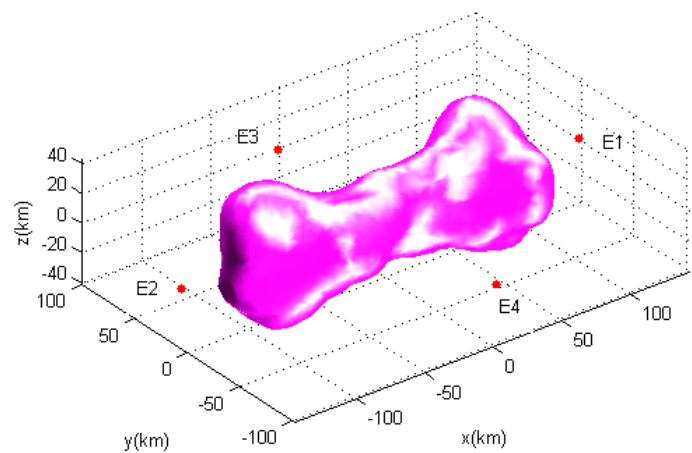

*Figure 1. Equilibrium points in the vicinity of asteroid 216 Kleopatra*

*Table 1. Positions of the equilibrium points around asteroid 216 Kleopatra (Jiang et al. 2014)*

| Equilibrium Points | x (km)   | y (km)   | z (km)    |
|--------------------|----------|----------|-----------|
| E1                 | 142. 852 | 2.45436  | 1.18008   |
| E2                 | -144.684 | 5.18855  | -0.282998 |
| E3                 | 2.21701  | -102.102 | 0.279703  |
| E4                 | -1.16396 | 100.738  | -0.541516 |

## 3. Periodic Orbits around Equilibrium Points

If the equilibrium point in the vicinity of an asteroid is linearly stable, there are three families of periodic orbits around the equilibrium point; if the equilibrium point is unstable, the number of periodic orbit families around the equilibrium point is less than three (Jiang et al. 2014).

### 3.1 Classification of Equilibrium Points

For the non-degenerate and non-resonant equilibrium points around an asteroid, the topological manifold classification (Jiang et al. 2014) of equilibrium points is presented in Figure 2. There are five cases for the non-degenerate and non-resonant equilibrium points in the potential field of a rotating asteroid where any two eigenvalues are unequal.

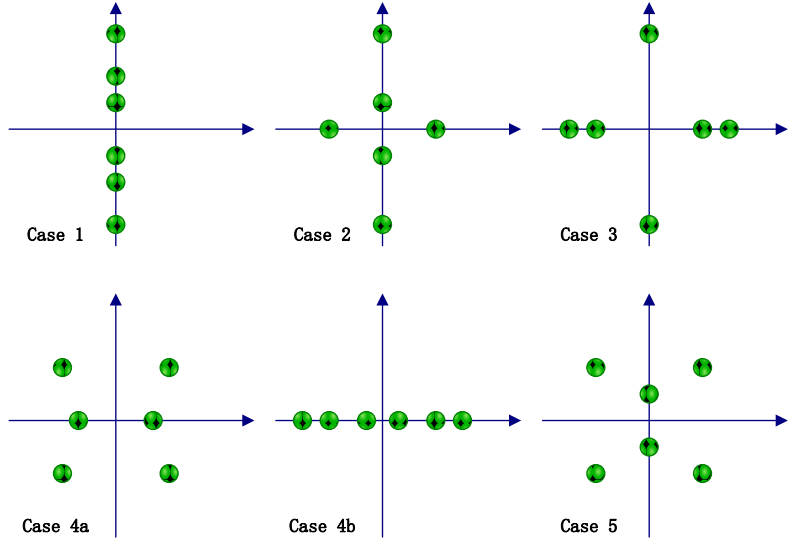

*Fig. 2. Classification of non-degenerate and non-resonant equilibrium points*

**Case 1**

In Case 1, the equilibrium point has three pairs of imaginary eigenvalues, and the equilibrium point is linearly stable. The linear motion around the equilibrium point is a quasi-periodic orbit, which is expressed as

$$\begin{bmatrix} \xi \\ \eta \\ \zeta \end{bmatrix} = \mathbf{C}_{3\times 6} \begin{bmatrix} \cos\beta_1 t & \sin\beta_1 t & \cos\beta_2 t & \sin\beta_2 t & \cos\beta_3 t & \sin\beta_3 t \end{bmatrix}^{\mathrm{T}}, \qquad (8)$$

where $\mathbf{C}_{3\times 6}$ is a $3\times 6$ real matrix.

**Case 2**

In Case 2, the equilibrium point has two pairs of imaginary eigenvalues and one pair of real eigenvalues, and the equilibrium point is unstable. The linear motion around the equilibrium point is composed of a quasi-periodic component, an asymptotically stable component and an unstable component, which are expressed as

$$\begin{bmatrix} \xi \\ \eta \\ \zeta \end{bmatrix} = \mathbf{C}_{3\times 6} \begin{bmatrix} e^{\alpha_1 t} & e^{-\alpha_1 t} & \cos\beta_1 t & \sin\beta_1 t & \cos\beta_2 t & \sin\beta_2 t \end{bmatrix}^{\mathrm{T}}. \qquad (9)$$

The asymptotically stable component $e^{-\alpha_1 t}$ is an exponential sink while the unstable component $e^{\alpha_1 t}$ is an exponential source, from which it follows that the projection of the equilibrium point in the asymptotically stable manifold is an exponential sink while in the unstable manifold the projection is an exponential source.

**Case 3**

In Case 3, the equilibrium point has one pair of imaginary eigenvalues and two pairs of real eigenvalues, and the equilibrium point is unstable. The linear motion around the equilibrium point is composed of a periodic component, two asymptotically stable components and two unstable components, which are expressed as

$$\begin{bmatrix} \xi \\ \eta \\ \zeta \end{bmatrix} = \mathbf{C}_{3\times 6} \begin{bmatrix} e^{\alpha_1 t} & e^{-\alpha_1 t} & e^{\alpha_2 t} & e^{-\alpha_2 t} & \cos\beta_1 t & \sin\beta_1 t \end{bmatrix}^{\mathrm{T}}. \qquad (10)$$

The asymptotically stable components $e^{-\alpha_1 t}$ and $e^{-\alpha_2 t}$ are exponential sinks while the unstable components $e^{\alpha_1 t}$ and $e^{\alpha_2 t}$ are exponential sources, similar to Case 2. The projection of the equilibrium point in the asymptotically stable manifold is an exponential sink while for the unstable manifold it is an exponential source.

**Case 4a**

In Case 4a, the equilibrium point has one pair of complex eigenvalues and one pair of real eigenvalues; the complex eigenvalues take the form $\pm\sigma\pm i\tau(\sigma,\tau\in\mathrm{R};\sigma,\tau>0)$. Hence the equilibrium point is unstable. The linear motion around the equilibrium point is composed of three asymptotically stable components and three unstable components, which are expressed as

$$\begin{bmatrix}\xi\\\eta\\\zeta\end{bmatrix}=\mathbf{C}_{3\times6}\begin{bmatrix}e^{\alpha_1 t} & e^{-\alpha_1 t} & e^{\sigma t}\cos\tau t & e^{\sigma t}\sin\tau t & e^{-\sigma t}\cos\tau t & e^{-\sigma t}\sin\tau t\end{bmatrix}^{\mathrm{T}}. \quad (11)$$

The asymptotically stable component $e^{-\alpha_1 t}$ is an exponential sink while the asymptotically stable components $e^{-\sigma t}\cos\tau t$ and $e^{-\sigma t}\sin\tau t$ are spiral sinks, meanwhile, the unstable component $e^{\alpha_1 t}$ is an exponential source while the unstable components $e^{\sigma t}\cos\tau t$ and $e^{\sigma t}\sin\tau t$ are spiral sources.

**Case 4b**

In Case 4b, the equilibrium point has three pairs of real eigenvalues and the equilibrium point is unstable. The linear motion around the equilibrium point is composed of three asymptotically stable components and three unstable components, which are expressed as

$$\begin{bmatrix}\xi\\\eta\\\zeta\end{bmatrix}=\mathbf{C}_{3\times6}\begin{bmatrix}e^{\alpha_1 t} & e^{-\alpha_1 t} & e^{\alpha_2 t} & e^{-\alpha_2 t} & e^{\alpha_3 t} & e^{-\alpha_3 t}\end{bmatrix}^{\mathrm{T}}. \quad (12)$$

The asymptotically stable component $e^{-\alpha_j t}\,(j=1,2,3)$ is an exponential sink while the unstable component $e^{\alpha_j t}\,(j=1,2,3)$ is an exponential source.

**Case 5**

In Case 5, the equilibrium point has one pair of imaginary eigenvalues and one pair of complex eigenvalues, and the equilibrium point is unstable. The linear motion around the equilibrium point is composed of a periodic component, two asymptotically stable components and two unstable components, which are expressed as

$$\begin{bmatrix} \xi \\ \eta \\ \zeta \end{bmatrix} = \mathbf{C}_{3\times 6} \begin{bmatrix} \cos\beta_1 t & \sin\beta_1 t & e^{\sigma t}\cos\tau t & e^{\sigma t}\sin\tau t & e^{-\sigma t}\cos\tau t & e^{-\sigma t}\sin\tau t \end{bmatrix}^{\mathrm{T}}. \quad (13)$$

The asymptotically stable components $e^{-\sigma t}\cos\tau t$ and $e^{-\sigma t}\sin\tau t$ are spiral sinks, while the unstable components $e^{\sigma t}\cos\tau t$ and $e^{\sigma t}\sin\tau t$ are spiral sources.

**3.2 Characteristic Multipliers of Orbits**

By solving Eq. (3) we can express the orbit $z$ as $\mathbf{z}(t) = \mathbf{f}(t, \mathbf{z}_0)$ where $\mathbf{f}(0, \mathbf{z}_0) = \mathbf{z}_0$. Let $S_p(T)$ be the set of periodic orbits with period $T$. Choose one periodic orbit from this set $p \in S_p(T)$, and consider the matrix of this periodic orbit $\nabla \mathbf{f} := \dfrac{\partial \mathbf{f}(\mathbf{z})}{\partial \mathbf{z}}$. It is a $6\times 6$ matrix. Therefore the state transition matrix is given by

$$\Phi(t) = \int_0^t \frac{\partial \mathbf{f}}{\partial \mathbf{z}}(p(\tau))d\tau. \quad (14)$$

Let $t = T$, then $\Phi(t = T)$ is the monodromy matrix for this periodic orbit. Denote the matrix as $M = \Phi(T)$. The eigenvalues of the monodromy matrix are the characteristic multipliers of the orbit (Marsden and Ratiu 1999). The state transition matrix $\Phi(t)$ is symplectic. If $\lambda$ is an eigenvalue of the matrix $\Phi(t)$, then $\lambda^{-1}$, $\bar{\lambda}$, and $\bar{\lambda}^{-1}$ are also eigenvalues of $\Phi(t)$.

## 3.3 Correspondence of the Topological Classification of Equilibrium Points and Topological Classification of Orbits

Suppose that eigenvalues of the equilibrium point take the form $\{\pm i\beta(\beta \in R^+), \lambda_3, \lambda_4, \lambda_5, \lambda_6\}$, then characteristic multipliers of periodic orbits around the equilibrium point take the form $\left\{1, 1, \exp\left(\frac{2\pi\lambda_j}{\beta}\right)(j=3,4,5,6)\right\}$ (Marsden and Ratiu 1999). The equilibrium point is resonant when at least two purely imaginary eigenvalues coincide. According to Jiang et al. (2014), the non-degenerate and non-resonant equilibrium points in the potential of an irregular celestial body can be classified into five different cases; this classification is shown in figure 2. Figure 3 shows the distribution of characteristic multipliers of orbits corresponding to the classification given by figure 2. The non-degenerate and non-resonant equilibrium point which has periodic orbits around it matches four of those five cases, which are Cases 1, 2, 3, and 5 in figure 2; periodic orbits around the equilibrium point also correspond to four cases: Cases 1, 2, 3, 5 in figure 3. So the distribution of characteristic multipliers of periodic orbits around the equilibrium point and the distribution of eigenvalues of the equilibrium point correspond to each other, and the distribution of eigenvalues of the equilibrium point confirms the various cases and stability of periodic orbits around the equilibrium point.

If the distribution of eigenvalues of the equilibrium point belongs to Case 1 in figure 2, then the distribution of characteristic multipliers of periodic orbits around the equilibrium point belongs to Case 1 in figure 3; i.e. if all the eigenvalues of the equilibrium point fall on the imaginary axis, the characteristic multipliers of periodic orbits around the equilibrium point are within the unit circle, and at least two of the characteristic multipliers are equal to 1. Table 2 shows the algebraic forms for the correspondence of eigenvalues of the equilibrium point and characteristic multipliers of periodic orbits around the equilibrium point.

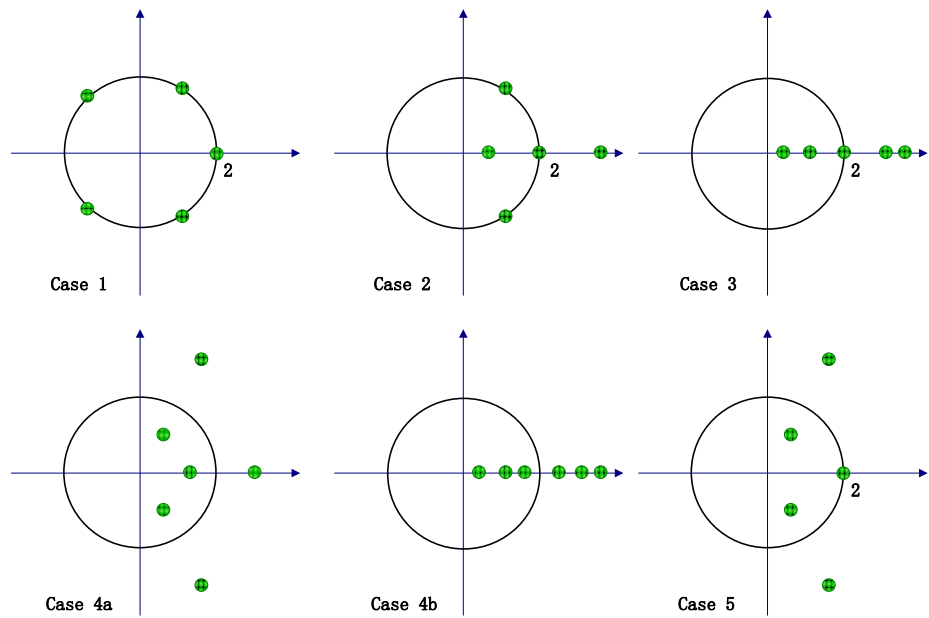

*Fig. 3. The distribution of characteristic multipliers for orbits corresponding to the classification in figure 2*

*Table 2. The correspondence of eigenvalues of the equilibrium point and characteristic multipliers of periodic orbits around the equilibrium point*

| Cases | eigenvalues of the equilibrium point | characteristic multipliers of periodic orbits around the equilibrium point |
|---|---|---|
| Case 1 | $\pm i\beta_j \begin{pmatrix} \beta_j \in \mathrm{R}, \beta_j > 0; j=1,2,3 \\ \forall k \neq j, k=1,2,3, s.t. \beta_k \neq \beta_j \end{pmatrix}$ | $\begin{cases} \tilde{\gamma}_j \left( \tilde{\gamma}_j = 1; j=1,2 \right) \\ e^{\pm i\tilde{\beta}_j} \left( \tilde{\beta}_j \in (0,\pi); j=1,2 \big| \tilde{\beta}_1 \neq \tilde{\beta}_2 \right) \end{cases}$ |

| Case 2 | $\begin{cases} \pm i\beta_j \left(\beta_j \in \mathrm{R}, \beta_j > 0; j=1,2 \mid \beta_1 \neq \beta_2\right) \\ \pm \alpha_j \left(\alpha_j \in \mathrm{R}, \alpha_j > 0, j=1\right) \end{cases}$ | $\begin{cases} \tilde{\gamma}_j \left(\tilde{\gamma}_j = 1; j=1,2\right) \\ e^{\pm i\tilde{\beta}_j} \left(\tilde{\beta}_j \in (0,\pi), j=1\right) \\ \mathrm{sgn}(\tilde{\alpha}_j) e^{\pm \tilde{\alpha}_j} \left(\tilde{\alpha}_j \in \mathrm{R}, |\tilde{\alpha}_j| \in (0,1), j=1\right) \end{cases}$ |
|---|---|---|
| Case 3 | $\begin{cases} \pm i\beta_j \left(\beta_j \in \mathrm{R}, \beta_j > 0; j=1\right) \\ \pm \alpha_j \left(\alpha_j \in \mathrm{R}, \alpha_j > 0; j=1,2 \mid \alpha_1 \neq \alpha_2\right) \end{cases}$ | $\begin{cases} \tilde{\gamma}_j \left(\tilde{\gamma}_j = 1; j=1,2\right) \\ \mathrm{sgn}(\tilde{\alpha}_j) e^{\pm \tilde{\alpha}_j} \begin{pmatrix} \tilde{\alpha}_j \in \mathrm{R}, |\tilde{\alpha}_j| \in (0,1); \\ j=1,2 \mid \tilde{\alpha}_1 \neq \tilde{\alpha}_2 \end{pmatrix} \end{cases}$ |
| Case 4a | $\begin{cases} \pm \sigma \pm i\tau \left(\sigma, \tau \in \mathrm{R}; \sigma, \tau > 0\right) \\ \pm \alpha_j \left(\alpha_j \in \mathrm{R}, \alpha_j > 0, j=1\right) \end{cases}$ | $\begin{cases} e^{\pm \tilde{\sigma} \pm i\tilde{\tau}} \left(\tilde{\sigma}, \tilde{\tau} \in \mathrm{R}; \tilde{\sigma} > 0, \tilde{\tau} \in (0,\pi)\right) \\ \mathrm{sgn}(\tilde{\alpha}_j) e^{\pm \tilde{\alpha}_j} \left(\tilde{\alpha}_j \in \mathrm{R}, |\tilde{\alpha}_j| \in (0,1), j=1\right) \end{cases}$ |
| Case 4b | $\pm \alpha_j \left(\alpha_j \in \mathrm{R}, \alpha_j > 0, j=1,2,3\right)$ | $\mathrm{sgn}(\tilde{\alpha}_j) e^{\pm \tilde{\alpha}_j} \left(\tilde{\alpha}_j \in \mathrm{R}, |\tilde{\alpha}_j| \in (0,1), j=1,2,3\right)$ |
| Case 5 | $\begin{cases} \pm i\beta_j \left(\beta_j \in \mathrm{R}, \beta_j > 0, j=1\right) \\ \pm \sigma \pm i\tau \left(\sigma, \tau \in \mathrm{R}; \sigma, \tau > 0\right) \end{cases}$ | $\begin{cases} \tilde{\gamma}_j \left(\tilde{\gamma}_j = 1; j=1,2\right) \\ e^{\pm \tilde{\sigma} \pm i\tilde{\tau}} \left(\tilde{\sigma}, \tilde{\tau} \in \mathrm{R}; \tilde{\sigma} > 0, \tilde{\tau} \in (0,\pi)\right) \end{cases}$ |

According to Jiang et al. (2014), the equilibrium point E1 around 216 Kleopatra belongs to Case 2. Table 3 shows eigenvalues of the equilibrium points around 216 Kleopatra from Jiang et al. (2014). There exist two families of periodic orbits around E1. Figure 4 presents the first family of periodic orbits corresponding to eigenvalues $\pm 0.413 \times 10^{-3}$i of E1, while figure 5 presents the second family of periodic orbits corresponding to eigenvalues $\pm 0.425 \times 10^{-3}$i of E1. Table 4 lists the periods of various periodic orbits in the first family around the equilibrium point E1 with the characteristic period 4.10569 h, and table 5 lists the periods of various periodic orbits in the second family around the equilibrium point E1 with the characteristic period 4.10569 h. In tables 4 and 5, the orbits with smaller index numbers correspond to the smaller orbits in figures 4 and 5, respectively. From tables 4 and 5 we see that the period of periodic orbits around the equilibrium point and the characteristic period of the equilibrium point are nearly the same.

*Table 3. Eigenvalues of the Equilibrium Points around Asteroid 216 Kleopatra (Jiang et al. 2014)*

| $\times 10^{-3} s^{-1}$ | $\lambda_1$ | $\lambda_2$ | $\lambda_3$ | $\lambda_4$ | $\lambda_5$ | $\lambda_6$ |
|---|---|---|---|---|---|---|
| E1 | 0.376 | -0.376 | 0.413i | -0.413i | 0.425i | -0.425i |
| E2 | 0.422 | -0.422 | 0.414i | -0.414i | 0.466i | -0.466i |
| E3 | 0.327i | -0.327i | 0.202+0.304i | 0.202-0.304i | -0.202+0.304i | -0.202-0.304i |
| E4 | 0.323i | -0.323i | 0.202+0.306i | 0.202-0.306i | -0.202+0.306i | -0.202-0.306i |

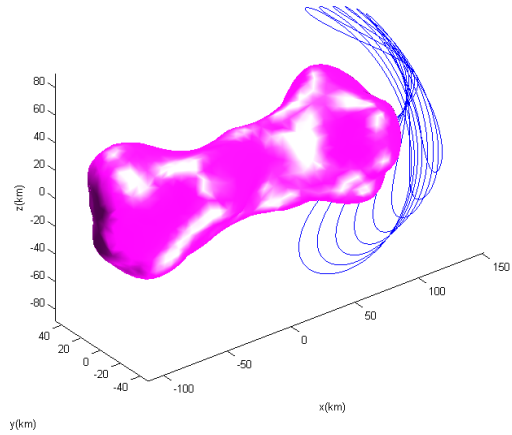

*Figure 4. The first family of periodic orbits around the equilibrium point E1 in the vicinity of asteroid 216 Kleopatra*

*Table 4. Period of different orbits in the first family around the equilibrium point E1 in the vicinity of asteroid 216 Kleopatra (characteristic period is 4.10569 h)*

| Orbits | T (h) |
|---|---|
| 1 | 4.18927 |
| 2 | 4.18145 |
| 3 | 4.18535 |
| 4 | 4.20040 |
| 5 | 4.22822 |
| 6 | 4.27043 |
| 7 | 4.32870 |
| 8 | 4.40425 |

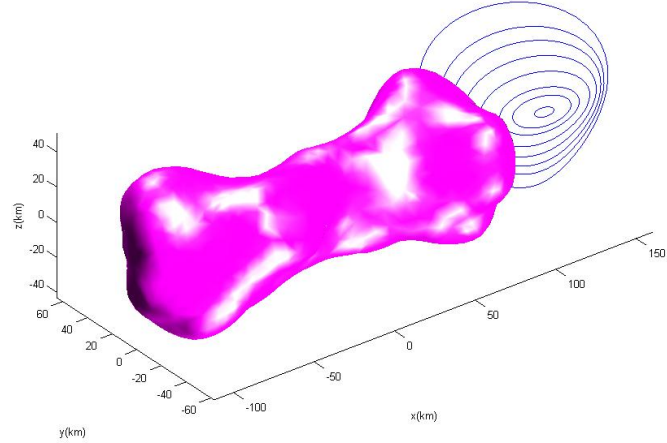

*Figure 5. The second family of periodic orbits around the equilibrium point E1 in the vicinity of asteroid 216 Kleopatra*

*Table 5. Period of different orbits in the second family around the equilibrium point E1 in the vicinity of asteroid 216 Kleopatra (characteristic period is 4.10569 h)*

| Orbits | T (h) |
|---|---|
| 1 | 4.10581 |
| 2 | 4.10309 |
| 3 | 4.09945 |
| 4 | 4.08941 |
| 5 | 4.07552 |
| 6 | 4.05795 |
| 7 | 4.03754 |
| 8 | 3.98919 |

The equilibrium points E1 and E3 belong to Case 2, while the equilibrium points E2 and E4 belong to Case 5. Thus there are a total of six families of periodic orbits in the vicinity of these four equilibrium points. We choose one periodic orbit from each family of periodic orbits and plot it in figure 6. Characteristic multipliers of these six periodic orbits are presented in table 6. Orbits 1 and 2 belong to E1, and have the same distribution of characteristic multipliers. From tables 3 and 6, it can be seen that the distribution of characteristic multipliers of orbits 1 and 2 corresponds to the distribution of eigenvalues of equilibrium point E1. Other orbits and equilibrium points show similar results. Therefore, we have shown that the distribution of characteristic multipliers of periodic orbits around the equilibrium point and the distribution of eigenvalues of equilibrium point correspond to each other.

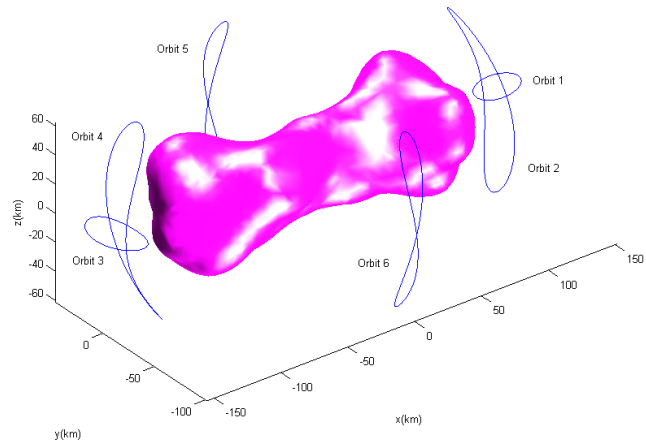

*Figure 6. Six periodic orbits around equilibrium points in the vicinity of asteroid 216 Kleopatra*

*Table 6. Characteristic multipliers of periodic orbits around equilibrium points in the vicinity of asteroid 216 Kleopatra*

|  | $\tilde{\lambda}_1$ | $\tilde{\lambda}_2$ | $\tilde{\lambda}_3$ | $\tilde{\lambda}_4$ | $\tilde{\lambda}_5$ | $\tilde{\lambda}_6$ |
|---|---|---|---|---|---|---|
| Orbit 1(E1) | 253.479 | 0.003945 | 0.98537+0.17041i | 0.98537-0.17041i | 1 | 1 |
| Orbit 2(E1) | 224.867 | 0.004447 | 0.96700+0.25478i | 0.96700-0.25478i | 1 | 1 |
| Orbit 3(E2) | 257.858 | 0.003878 | 0.76679+0.64189i | 0.76679-0.64189i | 1 | 1 |
| Orbit 4(E2) | 362.411 | 0.002759 | 0.76763+0.64090i | 0.76763-0.64090i | 1 | 1 |
| Orbit 5(E3) | 31.662+11.980i | 31.662-11.980i | 0.02763+0.01045i | 0.02763-0.01045i | 1 | 1 |
| Orbit 6(E4) | 35.122+ | 35.122- | 0.02695+ | 0.02695- | 1 | 1 |

|   | 8.3453i | 8.3453i | 0.00640i | 0.00640i |   |   |

## 4. Conclusions

We have studied the equilibrium points and periodic orbits in the potential of asteroids. The linearized equations and characteristic equations of motion relative to the equilibrium points are presented. The non-degenerate and non-resonant equilibrium point which has periodic orbits around it corresponds to four topological cases. We find that the distribution of characteristic multipliers of periodic orbits around the equilibrium point and the distribution of eigenvalues of equilibrium point correspond to each other. The distribution of eigenvalues of the equilibrium point confirms the topology and stability of periodic orbits around the equilibrium point. If all the eigenvalues of the equilibrium point are in the imaginary axis, the characteristic multipliers of periodic orbits around the equilibrium point are within the unit circle, and at least two of the characteristic multipliers are equal to 1.

**Acknowledgements**

This research was supported by the National Basic Research Program of China (973 Program, 2012CB720000), the State Key Laboratory Foundation of Astronautic Dynamics (No. 2015ADL0202), and the National Natural Science Foundation of China (No. 11372150).